# Multiple stable states of a periodically driven electron spin in a quantum dot using circularly polarized light

V.L. Korenev

A. F. Ioffe Physical Technical Institute, St. Petersburg, 194021 Russia

*The periodical modulation of circularly polarized light with a frequency close to the electron spin resonance frequency induces a sharp change of the single electron spin orientation. Hyperfine interaction provides a feedback, thus fixing the precession frequency of the electron spin in the external and the Overhauser field near the modulation frequency. The nuclear polarization is bidirectional and the electron-nuclear spin system (ENSS) possesses a few stable states. The same physics underlies the frequency-locking effect for two-color and mode-locked excitations. However, the pulsed excitation with mode locked laser brings about the multitudes of stable states in ENSS in a quantum dot. The resulting precession frequencies of the electron spin differ in these states by the multiple of the modulation frequency. Under such conditions ENSS represents a digital frequency converter with more than a hundred stable channels.*

PACS: 78.67.Hc 78.20.-e 78.47.-p

The electron spin precession in a quantum dot becomes disturbed as a result of hyperfine interaction with tens of thousands of surrounding nuclei. If nuclear spins are disordered, this leads to dephasing within some nanoseconds in a random hyperfine nuclear field [1]. Several ways of suppressing the random fields have been offered: (i) dynamic polarization of nuclear spins up to 100% [2]; (ii) suppression of fluctuations due to the strong feedback induced by hyperfine interaction [3]. The nuclear field locking effect with decreased fluctuations was observed under two-color CW excitation when the frequency difference of pump and probe lasers coincides with the Larmor spin precession frequency of the ground state electron [4]. It was explained by the hole-spin assisted dynamic nuclear polarization feedback process. The Larmor frequency locking was observed in quantum dots under periodic pulsed excitation [5], too: the spin precession of ensemble of electrons in transverse-to-beam magnetic field $\vec{B}$ is broken into a few modes of precession. The authors suggested a deep modulation of nuclear spin relaxation time: in these modes the optical excitation of QD is forbidden as a consequence of the optical pumping of electrons into the state noninteracting with light - "dark" state, and so the electron-nuclear spin dynamics gets frozen. The later results [6, 7], however, have shown an efficient nuclear polarization under resonant excitation of the optically forbidden transition, too. Ref. [8] shared the opinion of Ref.[5] but pointed to the useful role in the locking process of the effective magnetic field $\vec{B}_{eff}$ of the circularly polarized light along the laser beam (the optical Stark effect [9] or the inverse Faraday effect [10]). As a result, the optically induced frequency locking effect manifests itself in a variety of experiments and a few principally different explanations are proposed.

Here I show that a periodical modulation of the circularly polarized light close to the electron spin resonance (ESR) frequency in a transverse-to-beam magnetic field induces resonance changes of the electron spin orientation. In this optical analog of classical ESR the oscillating effective magnetic field $\vec{B}_{eff}$ of the circularly polarized light plays a role of the microwave field. The electron-nuclear hyperfine interaction provides a feedback, thus fixing the precession frequency of the electron spin in the external and the Overhauser field close to the modulation frequency. The nuclear polarization is bidirectional and possesses a few stable states. The same physics results in the frequency-locking effect for two-color and mode-locked excitations. The modulation of the circularly polarized light comes from the beating of the two fields for the former, and from the superposition of the multitudes of the narrow-band lasers for the latter one. A huge number of stable states appear in ENSS in a single quantum dot in the latter case. The resulting precession frequencies of the electron spin in the external and the Overhauser field differ in these states by the multiple of the modulation frequency. The mean electron spin value (which is otherwise a continuum variable) is also digitized. Under such conditions ENSS represents a digital frequency converter with more than a hundred stable channels instead of a few modes of precession as was previously thought.

The effect is based on the optical pumping of atoms in a transversal-to-beam magnetic field [11] and the optical Stark effect [9, 10]. Consider a resident electron of a quantum dot in a magnetic field $\vec{B} \| x$ with spin projection along $|\rightarrow\rangle$ and against $|\leftarrow\rangle$ the $x$-axis (Fig.1A,B). Let the quantum dot be illuminated along z-axis by a CW (continuous wave) circularly polarized light, and the laser frequency be detuned from the optical transition into the excited state of a negatively charged trion $|T-\rangle = |\rightarrow,\leftarrow\rangle(|\Uparrow\rangle - |\Downarrow\rangle)/\sqrt{2}$ as shown in Fig.1B. Two electrons form a singlet $|\rightarrow,\leftarrow\rangle$, while the state of a hole is described by the superposition of states $|\Uparrow\rangle(|\Downarrow\rangle)$ with momentum projection $+3/2(-3/2)$ onto the z-axis of structure growth. To make physics transparent I restrict myself by this 3-level system and do not consider another orthogonal excited state $|T+\rangle$ [12]. The circularly polarized light has two effects on the electron spin: (i) optical pumping into "dark state" (with mean spin $S_z = \mp 1/2$ for $\sigma^\pm$ helicity) [13] (ii) it acts on the electron spin as if it were an effective magnetic field $\vec{B}_{eff}$ along the laser beam [9, 10]. The spin precession frequency $\omega_{eff} \sim \rho_c \cdot \alpha^2 / \Delta$ about $\vec{B}_{eff}$ is determined by detuning frequency $\Delta$ from the optical resonance, the circular polarization degree $\rho_c$ of light and the Rabi frequency $\alpha$ of optical transition [14].

Assume that the laser circular polarization $\rho_c$ is periodically modulated with frequency $\omega$ close to the ESR precession frequency $\omega_x$ about the magnetic field, which is $\mu_B g_e B/\hbar$ in the absence of nuclear polarization ($\mu_B > 0$ is Bohr magneton, $g_e$ is the electron g-factor [15]). In a frame rotating about an external magnetic field $\vec{B}$ with frequency $\omega$ (Fig.1C) both the frequency $\vec{\omega}_{eff}$ and the injected initial spin polarization $\vec{S}_0$ are independent of time and directed along Z-axis (the rotating wave approximation). Z-axis coincides with z-axis in laboratory frame in the moments of $\sigma^+$ laser helicity with $S_0 = -1/2$. The electron mean spin $\vec{S}$ rotates about the vector $\vec{\omega}_L = \vec{\omega}_{eff} + (\vec{\omega}_x - \vec{\omega})$. In case of small decoherence spin $\vec{S}$ takes many turns, so that only the projection of $\vec{S}$ onto $\vec{\omega}_L$ is conserved. Similar to atomic systems [11] there is an optical spin orientation of QD electron along Z-axis

$$S_Z = S_0 \frac{\omega_{eff}^2}{(\omega_x - \omega)^2 + \omega_{eff}^2}$$

The Z-component of electron spin is maximal $S_Z = S_0 = -1/2$ in resonance ($\omega = \omega_x$). In this case at each period of Larmor precession, the angular momentum of light is added constructively to the precessing QD electron spin. In the laboratory frame (z, x) the mean electron spin rotates about the external field with a mean spin value being close to ½. However out of resonance $\omega \neq \omega_x$ the x-component of spin appears

$$S_x = S_0 \frac{(\omega_x - \omega) \cdot \omega_{eff}}{(\omega_x - \omega)^2 + \omega_{eff}^2} \tag{1}$$

which is time independent in the laboratory frame, too. Interestingly, the $S_x$ value may be very large (Fig.1D) even for the small $\omega_{eff}$ (provided that $\omega_{eff}$ exceeds characteristic electron spin relaxation rates). It is interesting to note that such a scheme can be considered as the optically induced electron spin resonance. Here the effective magnetic field $\vec{B}_{eff}$ of circularly polarized light plays a role of a microwave electromagnetic field. Periodical modulation of laser helicity (or the intensity) leads to the periodical oscillations of the $\vec{B}_{eff}$-field and to ESR signal without an external microwave field.

Earlier the optically induced nuclear magnetic resonance (NMR) was discovered in semiconductors [16] (called recently "all-optical NMR").

The quantum dot electron undergoes a hyperfine interaction with huge number of surrounding nuclear spins $N \sim 10^4$. In nonequilibrium conditions this leads to the dynamic polarization of nuclei parallel to the external magnetic field [13]. The contact part of hyperfine interaction enables mutual flipping of electron and nuclear spins – flip-flop transitions transferring the $S_x$ electron spin to nuclei and inducing a non-zero mean nuclear spin $I_N$. In their turn, the polarized nuclei create a hyperfine Overhauser magnetic field, which affects the precession frequency of the electron spin. As a result, there appears a non-linear ENSS, in which the electron spin polarizes nuclei and depends itself on their mean spin. Now the frequency $\omega_x = \omega_{Zeeman} + \omega_N$ in Eq.(1) is the sum of precession frequencies $\omega_{Zeeman} = \mu_B g_e B / \hbar$ in the external and $\omega_N = A I_N / \hbar$ ($A$ - hyperfine constant) in Overhauser fields. The dynamic polarization follows two ways in quantum dots [17]: (i) the Overhauser effect [18], taking place when the distribution of electrons over spin levels does not correspond to the lattice temperature that provides energy for flip-flop transitions; (ii) solid effect [18], recently named "reverse Overhauser effect". Here the provider of the energy for flip-flop is the external field [19], rather than the lattice. The solid effect is essential in the case of resonant excitation of quantum dots, when the charge carriers are strongly localized and isolated from the lattice. The optical field provides the energy. This is the optical version of the solid-effect. The equation of dynamic nuclear polarization via the "optical solid effect" is

$$\frac{dI_N}{dt} = -\frac{1}{T_{1e}}[I_N - QS_x(I_N)] - \frac{1}{T_{1N}}I_N \qquad (2)$$

where $Q = 4I(I+1)/3$, the nuclear polarization time $T_{1e}$ and the spin-lattice relaxation time $T_{1N}$ take into account possible leakage [20]. In the stationary case one transcendental equation following from Eqs.(1,2) determines the nuclear spin $I_N$ (Appendix C). Two parameters $a = A/\hbar|\omega_{eff}|$ and $b = f_N Q S_0 \vec{\omega}_{eff} / |\omega_{eff}|$ enter this problem, where the leakage factor $f_N = T_{1N}/(T_{1e} + T_{1N})$. Figure 2A shows the nuclear spin $I_N$ vs detuning $\omega_{Zeeman} - \omega$ from the unperturbed (by nuclear spin orientation) resonance under $a = 200, b = -0.5$. As it follows from the dynamic Eq. (2) the stationary points with a slope of the function $f_N Q \partial S_x(I_N)/\partial I_N < 1$ are stable. Solid (dashed) lines in Fig2A correspond to the stable (unstable) states. Nuclear spin is bi-directional following the electron spin polarization $S_x$. It is seen in Fig.2B that in a wide range of detuning $\omega_{Zeeman} - \omega$ the nuclear spin adjusts itself to keep the spin precession frequency $\omega_x$ very close (i.e. locked) to modulation frequency $\omega$. Near each edge of the plato there is bistability, so that the hysteretic phenomena are expected. Also there is enhanced spin orientation of electron $(S_Z \sim S_0)$ in the wide range of detuning within the plato. It will manifest itself in sizable oscillations of electron spin $S_z$ component in laboratory frame, which can be detected by the Kerr rotation technique. The plato is stable for the positive detuning $\Delta$ from the optical resonance (Fig.1B). The $\Delta$ sign change inverts both the $\vec{\omega}_{eff}$ and $S_x$ directions (Eq.1). Although the bistability regions still exist, the $\omega_x$ locking is absent (the corresponding states are unstable).

The same model predicts the frequency locking effect in the case of two-color CW and mode-locked (pulsed) excitations of a single QD. The two-color excitation of a quantum dot by two CW lasers with $\Omega_1, \Omega_2$ frequencies (at $\Omega_2 - \Omega_1 \approx \omega_x$) and orthogonal constant in time linear polarizations is usually described in terms of coherent population trapping (optical pumping of "dark" state) and 3-level $\Lambda$–scheme [21, 22]. Actually it is very similar to the foregoing consideration, because the superposition of such two electromagnetic waves gives a wave whose circular polarization oscillates on single frequency $\omega = \Omega_2 - \Omega_1$. Indeed, the electric field $\vec{E}(t)$ component of two linearly polarized electromagnetic waves of the same amplitude $E$ is

$$\vec{E}(t) = \frac{E}{2}\vec{e}_1 \exp(i\Omega_1 t) + \frac{E}{2}\vec{e}_2 \exp(i\Omega_2 t + i\Phi) + c.c.$$

where $\vec{e}_1(\vec{e}_2)$ is the unit vector along x (y) axis in laboratory frame, $\Phi$ is the constant phase shift. The degree of circular polarization of light (Stokes parameter) $\rho_c = \sin(\tilde{\Phi})$ is determined by the total phase difference $\tilde{\Phi}(t) = (\Omega_2 - \Omega_1) \cdot t + \Phi$. The degree $\rho_c(t)$ oscillates from +1 ($\sigma^+$-light) to -1 ($\sigma^-$-light) at a single frequency $\omega = |\Omega_2 - \Omega_1|$. In this sense the two-color excitation is equivalent to the previous discussion of CW laser with modulated helicity, which is not a single frequency light source anymore. Hence the same Eq.(1) (and Fig.1C) holds for the two-color excitation. It is derived in Appendix A from the matrix of density solution of the $\Lambda$-scheme (Fig.1B) according to Ref.[23]. The hyperfine interaction with electrons locks the frequency $\omega_x = \omega_{Zeeman} + \omega_N$ and bidirectional nuclear polarization takes place as discussed above. This conclusion is confirmed qualitatively by the recent two-laser experiment [4] reporting the nuclear field locking.

In case of excitation by pulsed laser in the mode-locking regime [5, 8, 24] the amplitude of the electromagnetic field is a superposition of a great number of monochromatic waves, whose frequencies are evenly distributed around the carrier frequency. For example, at pulse duration $1\,ps$ and periodicity $T_R = 2\pi/\Omega_R = 13\,ns$ more than *10000* coherent waves excite the quantum dot simultaneously. Time evolution of both $S_0(t)$ and $\omega_{eff}(t)$ consists of many harmonics multiple to $\Omega_R$. The optical orientation of the electron spin takes place when the difference of frequencies (multiple to $\Omega_R$) of any two waves is close to the synchronization condition

$$K \cdot \Omega_R = \omega_x \qquad (3)$$

where *K* is positive integer. The spectral dependence of $S_x$ can be obtained (Appendix B) according to Ref.[24] developed for the case of optical Stark effect (the relevant entity here is the rotation angle $\varphi$ about z-axis during pulse duration). The blue curve in Fig.2C shows the fragment of the dependence of $S_x$ on precession frequency $\omega_N$ in the nuclear field in $\Omega_R$ units at $\varphi = 0.1$ and $\omega_{Zeeman}/\Omega_R = 1000$ (the specific value of the latter is not of much importance because it just shifts the comb in Fig.2C along the abscissa axis). Physically this curve can be imagined as the replication of the Fig.1D every K-th mode. According to Eq.(2) the stationary states are given by the crossing points of $S_x(\omega_N)$ with the line $I_N/Qf_N$ (Fig.2C). There is a huge number of stable modes in a single dot – ideally about one thousand. Indeed, the frequency $\omega_x = \omega_{Zeeman} + \omega_N$ involves the precession frequency in Overhauser field $\omega_N = AI_N/\hbar$, while the mean nuclear spin $I_N$ may vary from $-I$ to $I$. The electron-nuclear coupling feedback brings about a new stable state of ENSS, in which the precession frequency of the electron spin is near condition (3). Tuning up to the next resonance is carried out through the self-adjustment of the frequency $\omega_N$. Resonances are distant from each other by a magnitude multiple to $\Omega_R$, therefore the maximal number of resonances overlapped by nuclei is $2AI/\hbar\Omega_R \approx 840$ (at $A = 90\,\mu eV$, $I = 3/2$ [13], $\Omega_R = 0.5\,ns^{-1}$ [24]). Thus the Overhauser field in a quantum dot becomes "digitalized" [25], so that frequency $\omega_x$ takes one of 840 stable values, and the quantum dot works as a digital converter of Larmor frequency [26]. One can see that the mean electron spin $S_x$ value (which is otherwise is a continuum variable) is also digitized. In this specific example the corresponding steps are $\Delta S_x \approx \hbar\Omega_R/AQf_N \approx 10^{-2} \div 10^{-3}$. In turn, the mean nuclear spin $I_N$ falls in between $\pm Q \cdot f_N \cdot S_0/2 = \pm 0.25$, so the number of states mounts up to 140. In each pair of crossing points in Fig.2D only solid points are stable (see also solid points in Fig.2C). What specific ENSS state is realized depends on initial conditions (the initial value of nuclear polarization). Periodical excitation locks the precession frequency to the stable state nearest to the starting frequency value [27].

Spin initialization is an important condition of any spin-based program in quantum information area. Here each member of the QD ensemble can be labeled by the specific precession frequency. The simplest way to mark the dots is to illuminate first the QD ensemble by the CW laser modulated at single frequency. After one minute (typical nuclear polarization time upon the light [5]) the nuclear spin system of QD ensemble becomes polarized along the tilted straight line in Fig.2A (those out of the line are not optically oriented and can be easily excluded [28]). The mean nuclear polarization $I_N$ varies from dot to dot because of non-uniform broadening in the relative detuning from the electron spin resonance (scattering over the abscissa axis in Fig.2A). As a second step, one switch on the pulsed laser (the modulated CW laser is switched off), which locks $I_N$ distribution into the discrete number (>100) of the stable states (nearest to the initial values) as shown in Fig.2C. In each of these states the ESR frequency is a multiple of $\Omega_R$. This completes the preparation of the QD ensemble. The polarization memory without illumination can be stored for one hour [5].

This approach explains naturally the frequency focusing in a quantum dot ensemble in the pulsed regime without invoking the hypothesis [5] of freezing of nuclear dynamics. The stable points with $\omega_x \approx K\Omega_R$ are realized for the dots with positive detuning $\Delta$ from optical resonance as we discussed above. The in-phase spin precession of the optically oriented QD electrons increases the pump-probe signal in these dots. However, the locking effect is absent for the dots with $\Delta < 0$: these states are unstable because the sign of $\omega_{eff}$ is reversed and the slope of the function $S_x(I_N)$ in the stationary point is much larger than unity. The stable points are close to $(K+1/2)\Omega_R$ [29]. In these stable points the optical orientation of QD electrons is absent. Such dots give minor contribution into the pump-probe signal.

Hence the periodical modulation of circularly polarized light with a single frequency initiates the ESR optically without any external microwave field. Hyperfine interaction provides a feedback, thus fixing the precession frequency of the electron spin in the external and the Overhauser field close to the modulation frequency. The nuclear polarization is bidirectional and the ENSS possesses a few stable points. In turn, pulsed excitation with mode locked laser brings about the appearance of huge number of stable states in ENSS in a quantum dot. The resulting precession frequencies of the electron spin in the external and the Overhauser field differ in these states by the multiple of the modulation frequency. Under such conditions ENSS represents a digital frequency converter with more than a hundred stable channels. This opens the route for controllable preparation of the initially disordered ensemble of ~100 quantum dots. Each of the dots can be labeled by a fixed Larmor frequency.

I am grateful to I.A. Akimov, K.V. Kavokin, D.R. Yakovlev for fruitful discussions. This work was supported by DFG 436RUS113/958/0-1, RFBR, Program of RAS.

**APPENDIX A: Derivation of the Eq.(1) for the case of two-laser coherent excitation.**

Here I exploit the matrix of density $\hat{\rho}$ approach for the case of the two-color coherent excitation to show its equivalence to the case of single frequency modulation of circular polarization of light and to derive the Eq. (1) of the main text. Let the two orthogonally linearly polarized lasers of the same intensity with frequencies $\Omega_1$ and $\Omega_2$ induce $1 \leftrightarrow 3$ and $2 \leftrightarrow 3$ transitions as shown in $\Lambda$ – scheme in Fig.3. The $\Lambda$ – scheme was considered generally in Ref.[23] for the atoms to discuss the electromagnetically induced transparency (depletion of state 3). I adopt this approach to the QD case and focus on the optical orientation of the ground state electron. The states 1 and 2 correspond to the electron spin projections along $|\rightarrow\rangle$ and opposite $|\leftarrow\rangle$ to the $x$ – axis (Fig.1A,B), whereas the state 3 – to the trion state $|T-\rangle$. The $1 \leftrightarrow 3$ and $2 \leftrightarrow 3$ transitions are optically active in two orthogonal $x$ and $y$ linear polarizations, respectively [4], with the same absolute values of dipole matrix elements $|\mu_{13}| = |\mu_{23}|$. The generalized Rabi frequencies $\alpha = \frac{\mu_{13} E}{2\hbar}$ and $\beta = \frac{\mu_{23} E}{2\hbar} e^{i\Phi}$ of the optical transitions are assumed to be the same for both (this can be done by choosing an appropriate phase shift $\Phi$ between the two waves) and real (the absolute phase is not important), i.e. $\alpha = \beta = \alpha^* = \beta^*$. In the rotating wave approximation a simplified system of the equations of Ref. [23] reads

$$\dot{\rho}_{11} = i\alpha(\tilde{\rho}_{31} - \tilde{\rho}_{13}) + \frac{\rho_{33}}{T_2} \tag{A1a}$$

$$\dot{\rho}_{22} = i\alpha(\tilde{\rho}_{32} - \tilde{\rho}_{23}) + \frac{\rho_{33}}{T_2} \tag{A1b}$$

$$\rho_{11} + \rho_{22} + \rho_{33} = 1 \tag{A1c}$$

$$\dot{\tilde{\rho}}_{13} + i\left(\Delta - \frac{i}{T_2}\right)\tilde{\rho}_{13} = i\alpha(\rho_{33} - \rho_{11}) - i\alpha\tilde{\rho}_{12} \tag{A1d}$$

$$\dot{\tilde{\rho}}_{23} + i\left(\Delta' - \frac{i}{T_2}\right)\tilde{\rho}_{23} = i\alpha(\rho_{33} - \rho_{22}) - i\alpha\tilde{\rho}_{21} \tag{A1e}$$

$$\dot{\tilde{\rho}}_{12} + i\left(\delta - \frac{i}{\tau_2}\right)\tilde{\rho}_{12} = i\alpha(\tilde{\rho}_{32} - \tilde{\rho}_{13}) \tag{A1f}$$

where the decoherence time $T_2$ of excited (trion) state is assumed to be determined by the radiative decay with time $T_1$ (in the notation of [23]) of the excited state, so that $T_2 = 2T_1$. The decoherence time $\tau_2$ of the ground state (electron) is much longer than that of excited state, i.e. $\tau_2 \gg T_2$. The magnetic field suppresses strongly the spin-flip transitions between the ground states 1 and 2, so I removed the longitudinal electron spin relaxation at all: $\tau_1 = \infty$. Laser frequencies $\Omega_1(\Omega_2)$ of the waves and transition frequencies $\omega_{31}, \omega_{32}, \omega_{12}$ are shown in Fig.3. Parameters $\Delta = \Omega_1 - \omega_{31}$, $\Delta' = \Omega_2 - \omega_{32}$ and $\delta = \Delta - \Delta' = \Omega_1 - \Omega_2 + \omega_{12}$ denote the detuning from the two optical resonances $\Delta(\Delta')$ and from the two-photon Raman resonance $\delta$, which I call below the microwave resonance. Rapidly oscillating non-diagonal terms are removed with the definitions

$$\rho_{13} = \tilde{\rho}_{13}\exp(i\Omega_1 t) \tag{A2a}$$

$$\rho_{23} = \tilde{\rho}_{23}\exp(i\Omega_2 t) \tag{A2b}$$

$$\rho_{12} = \tilde{\rho}_{12}\exp[i(\Omega_1 - \Omega_2)t] \tag{A2c}$$

where spatial dispersion is omitted. Equations A1(a-c) reflect the population dynamics with the Eq. (A1c) expressing the conservation of charge equal to single electronic charge. Equations A1(d-f) describe the kinetics of the optical and ground state coherences. Although the system A1(a-e) looks much simpler than that considered in Ref.[23], it involves all the important physics: optical pumping, optical Stark effect and the electromagnetically induced transparency. I will not consider the latter effect here and refer to Ref.[22]. Further simplification can be achieved if we restrict ourselves by the low power regime ($\alpha T_2 \ll 1$), when the population of excited state is negligible, i.e. $\rho_{33} \ll \rho_{11}, \rho_{22}$. In this case the 3-level system becomes effectively two-level: the Equations A1(d,e) shortly ($\sim T_2$) reach the steady state. Thus we have for the ground state electron

$$(\dot{\rho}_{11} - \dot{\rho}_{22}) = 2\alpha \, \text{Im}(\tilde{\rho}_{13} + \tilde{\rho}_{32}) \tag{A3a}$$

$$\dot{\tilde{\rho}}_{12} + i\left(\delta - \frac{i}{\tau_2}\right)\tilde{\rho}_{12} = i\alpha(\tilde{\rho}_{32} - \tilde{\rho}_{13}) \tag{A3b}$$

$$\rho_{11} + \rho_{22} = 1 \tag{A3c}$$

where

$$\tilde{\rho}_{13} = -\alpha \frac{\rho_{11} + \tilde{\rho}_{12}}{\Delta - i/T_2} \, ; \; \tilde{\rho}_{32} = -\alpha \frac{\rho_{22} + \tilde{\rho}_{12}}{\Delta' + i/T_2} \tag{A3d}$$

We are interested in the effects, which happen close to the microwave resonance $\delta = \Delta - \Delta' \geq 1/\tau_2$. The optical resonance is smeared because $T_2 \ll \tau_2$. It means that $\delta T_2 = (\Delta - \Delta')T_2 \ll 1$, so one can put $\Delta = \Delta'$ in Eqs.(A3d). However the dimensionless parameter $\Delta T_2$ can take any value. Then putting Eqs.(A3d) into Eqs. A3(a-b) one obtains after some algebra a set of dynamic equations for the matrix elements $\rho_{11} - \rho_{22}, \tilde{\rho}_{12}$. It is convenient to express these elements through the electron spin components $S_{\tilde{x}}, S_{\tilde{y}}, S_{\tilde{z}}$ using the relations:

$$S_{\tilde{x}} = \frac{1}{2}Sp(\sigma_x \hat{\rho}) = \text{Re}\,\tilde{\rho}_{12}; \; S_{\tilde{y}} = \frac{1}{2}Sp(\sigma_y \hat{\rho}) = -\text{Im}\,\tilde{\rho}_{12}; \; S_{\tilde{z}} = \frac{1}{2}Sp(\sigma_z \hat{\rho}) = \frac{\rho_{11} - \rho_{22}}{2}$$

where trace (Sp) is performed over the product of $2 \times 2$ matrix of density of electron with Pauli matrices $\sigma_x, \sigma_y, \sigma_z$. The dynamic equations for the mean spin read as

$$\dot{S}_{\tilde{x}} = -\frac{1/2 + S_{\tilde{x}}}{\tau_J} - \frac{S_{\tilde{x}}}{\tau_2} - \omega_{\tilde{z}} S_{\tilde{y}} \tag{A4a}$$

$$\dot{S}_{\tilde{y}} = -\frac{S_{\tilde{y}}}{\tau_J} - \frac{S_{\tilde{y}}}{\tau_2} + \omega_{\tilde{z}} S_{\tilde{x}} - \omega_{\tilde{x}} S_{\tilde{z}} \tag{A4b}$$

$$\dot{S}_{\tilde{z}} = -\frac{S_{\tilde{z}}}{\tau_J} + \omega_{\tilde{x}} S_{\tilde{y}} \tag{A4c}$$

where $\dfrac{1}{\tau_J} = \dfrac{2\alpha^2 T_2}{1+(\Delta T_2)^2}$, $\omega_{\tilde{z}} = \delta$ and $\omega_{\tilde{x}} = \dfrac{2\alpha^2 \Delta T_2^2}{1+(\Delta T_2)^2}$. The physics is straightforward. The system of Bloch equations (A4) describes the optical pumping of electron spin in a rotating frame into the state with momentum projection opposite to $\tilde{x}$ – axis with characteristic time $\tau_J$. Additionally, the $\tilde{x}, \tilde{y}$ spin components undergo decoherence with time $\tau_2$. Besides these processes there is spin rotation due to the torque $\vec{\omega}_L \times \vec{S}$ with Larmor frequency $\vec{\omega}_L = (\omega_{\tilde{x}}, 0, \omega_{\tilde{z}})$ in $\{\tilde{x}, \tilde{y}, \tilde{z}\}$ frame. Thus we have one-to-one correspondence between the case of polarization-modulated excitation with single frequency $\omega$ (Fig.1C) and the excitation by two monochromatic waves with constant polarizations with frequency difference $\Omega_2 - \Omega_1$. The $(\tilde{x}, \tilde{y}, \tilde{z})$ spin components represent the mean electron spin $\vec{S}$ in a frame rotating about the $\tilde{z}$ – axis with frequency $\omega = \Omega_2 - \Omega_1$, as it follows from the Eq.(A2c). We identify the $\tilde{z}$ – axis ($\tilde{x}$ – axis) with x-axis (Z-axis) in Fig.1C of the main text. The frequency $\omega_{\tilde{z}} = \omega_{12} + \Omega_1 - \Omega_2$ is nothing but $\omega_x - \omega$: the electron spin splitting $\hbar\omega_x = \hbar\omega_{21}$, frequency $\omega = \Omega_2 - \Omega_1$. The $\omega_{\tilde{x}}$ component, which is the $\omega_{\text{eff}}$ in Fig.1C, describes the optical Stark effect (or the inverse Faraday effect): the effective magnetic field $B_{\text{eff}}$ is constant in the rotating frame and is periodically modulated in time at frequency $\Omega_2 - \Omega_1$ in laboratory frame. It should be noted that the expression for the frequency $\omega_{\text{eff}}$

$$\omega_{\text{eff}} = \dfrac{2\alpha^2 \Delta T_2^2}{1+(\Delta T_2)^2}$$

is valid in the whole range of detuning $\Delta$ from the optical resonance. In a particular case of large detuning $\Delta T_2 \gg 1$ we obtain the result of perturbation theory: $\omega_{\tilde{x}} = \omega_{\text{eff}} \approx \dfrac{2\alpha^2}{\Delta}$. In these conditions parameter $\omega_{\text{eff}} \tau_J = \Delta T_2$ is very large, so the electron spin makes a lot of turns (provided that the ground state decoherence is also long enough, $\omega_{eff}\tau_2 >> 1$) about the vector $\vec{\omega}_L$ before renewal/decoherence process. As a result only the projection of electron spin onto $\vec{\omega}_L$ is conserved

$$\vec{S} = \frac{(\vec{S}_0 \vec{\omega}_L)\vec{\omega}_L}{\omega_L^2} \tag{A5}$$

Hence the Z and x components of Eq. (A5) are

$$S_Z = S_0 \frac{\omega_{eff}^2}{(\omega_x - \omega)^2 + \omega_{eff}^2}, \quad S_x = S_0 \frac{(\omega_x - \omega) \cdot \omega_{eff}}{(\omega_x - \omega)^2 + \omega_{eff}^2} \tag{A6}$$

The first formula in Eq. (A6) describes the amplitude of Z-component $S_Z$ of the electron spin in the rotating frame. The Z-component reaches maximum in resonance $\omega = \omega_x$. It rotates in laboratory frame with the frequency $\omega = \Omega_2 - \Omega_1$. The second formula gives the projection of the electron spin onto external field. It is the same in rotating and laboratory frames. It is the Eq.1 of the main text.

**APPENDIX B. Excitation with a short-pulse train**

Here the spectral dependence of $S_x$ (blue curve in Fig2C) is calculated. The excitation by short pulses (shorter than the trion decay time $T_1$ and the spin precession times about the external magnetic field) of circularly polarized light induces the optical pumping in a high transversal magnetic field ($\mu_B g B T_1/\hbar \gg 1$) during the pump pulse $\tau_p$ [24]. In turn, the optically induced rotation of electron spin (the relevant entity here is the rotation angle $\varphi$ about the light beam) also happens during $\tau_p$ time. The resulting action of circularly polarized n-th pulse on the electron spin at a low power density can be described by equation

$$\vec{S}_n^+ - \vec{S}_n^- = g(\vec{S}_0 - \vec{S}_n^-) + \vec{\varphi} \times \vec{S}_n^- \tag{B1}$$

where $\vec{S}_n^-(\vec{S}_n^+)$ is the mean electron spin before (after) pulse arrival, $g$ – parameter characterizing the efficiency of optical pumping into the state $\vec{S}_0 = (0, 0, -1/2)$ (the spin components are in laboratory frame x,y,z). The meaning of Eq.(B1) is following. We have shown that under two-color optical excitation with low power density the 3-level system becomes effectively two-level and the evolution of the electron spin $\vec{S}(t)$ is described by the Bloch equations for the electron spin. In the case of pulsed excitation it is reasonable to replace the derivative term $\dot{\vec{S}}$ of Bloch equations with a difference $\vec{S}_n^+ - \vec{S}_n^-$. The first term in the RHS of (B1) describes the optical pumping similar to $\tau_J$ – terms in Eqs.(A4). The second term in the RHS of (B1) gives the optical Stark effect and is similar to $\omega_{\tilde{x}}$ – terms in Eqs.(A4). It describes the rotation of electron spin about the effective pulsed magnetic field directed along the beam. Eq.(B1) ignores both decoherence and the rotation about the external magnetic field during the pulse, because the pulse is assumed to be very short. The parameter $g$ is determined by the pulsed area $\theta = \int \alpha(t) dt$ instead of the Rabi frequency.

Similar to the previous analysis (Appendix A) one can expect that the rotation angle $\varphi \sim \rho_c \dfrac{\int \alpha^2(t)dt}{\Delta}$ where the detuning $\Delta$ is determined by the shift of the center of gravity of the pulse spectrum from the optical resonance. The angle $\varphi$ changes sign with $\Delta$ inversion and possesses a dispersion-like $\Delta$–dependence. In contrast to it the optical pumping parameter $g$ is related with absorption, hence $g \sim \dfrac{\theta^2}{\Delta^2}$ at large detuning $\Delta\tau_p \gg 1$. Therefore one can expect that the optical Stark effect dominates over the pumping rate $\varphi > g$ at large detuning (similar to what we had in the Appendix A $\omega_{eff}\tau_J \gg 1$ under $\Delta T_2 \gg 1$). In another particular case of exact optical resonance $\Delta = 0$ the angle $\varphi = 0$, whereas the pumping rate is maximal. Eq.(B1) in this case coincides with that derived in SOM of Ref.[24] with $g = \dfrac{\theta^2}{8}$. Below I consider a large detuning case when the optical Stark effect dominates, i.e. $\varphi \gg g$.

The spin polarization $\vec{S}(t)$ between n-th and n+1-th pulses evolves in laboratory frame x,y,z as

$$S_x(t) = S^+_{n,x};\quad S_y(t) = S^+_{n,y}\cos(\omega_x t) - S^+_{n,z}\sin(\omega_x t);\quad S_z(t) = S^+_{n,z}\cos(\omega_x t) + S^+_{n,y}\sin(\omega_x t) \quad (B2)$$

where the magnetic field is along x-axis, the light beam is along z-axis. The decoherence during the time $T_R$ between pulses is neglected. In the steady state regime the mean spin before n-th pulse is equal to that before n+1-th pulse:

$$\vec{S}^-_n = \vec{S}^-_{n+1} \equiv \vec{S}(\omega_x) \quad (B3)$$

The Eqs.(B1-B3) enable one to find $\vec{S}(\omega_x)$. The projection of spin onto magnetic field direction for the case $\varphi \gg g$ is given by

$$S_x(\omega_x) = \dfrac{1}{2}\dfrac{\sin(\omega_x T_R)\sin(\varphi)}{\cos(\omega_x T_R) + \cos(\varphi) + \cos(\varphi)\cdot\cos(\omega_x T_R) - 3} \quad (B4)$$

The spectral dependence (B4) is plotted in Fig.2C under $\varphi = 0.1$. One can see that Eq.(B4) represents a periodically repeated dispersion-like curves centered at the Larmor precession frequency $\omega_x$ multiple of $\Omega_R$. They can be interpreted as the optical pumping of electron spin in a frame rotating with frequency $\omega = K\Omega_R \approx \omega_x$.

In spite of the fact that the dependence in Fig.2C is obtained for the particular case of low power density ($\varphi \ll 1$) and large detuning ($g \ll \varphi$) the conclusion about many states in ENSS is quite general. The fact is that the periodicity of $S_x(\omega_x) = S_x(\omega_x + \Omega_R)$ is a general property of this periodically driven system, therefore the number of crossing points remains large. The specific set of parameters $g, \varphi$ determines only the specific shape of the $S_x(\omega_x)$ curve.

Near the K-th synchronization condition $\omega_x = K\Omega_R$ the Eq.(B4) looks very similar to the Eq.(1) of the main text (written for the modulation at single frequency $\omega$)

$$S_x(\omega_x) \approx -\frac{1}{2}\frac{(\omega_x - K\Omega_R)\cdot(\varphi/T_R)}{(\omega_x - K\Omega_R)^2 + (\varphi/T_R)^2} \tag{B5}$$

where the quantity $\varphi/T_R$ has a meaning of the $\omega_{eff}$ in Eq.(1). This expresses once more an inherent relation between the modulation at single frequency and the periodic pulse train: in both cases there is the optical pumping of the electron spin in rotating frame near the synchronization condition.

**APPENDIX C: Non-linear properties of the ENSS for polarization modulation at single frequency.**

Here I derive the steady state equations for the nuclear mean spin $I_N$ and the spin precession frequency $\omega_x = \omega_{Zeeman} + AI_N/\hbar$ in total external and Overhauser fields as a function of detuning $\omega_{Zeeman} - \omega$ from the non-perturbed microwave resonance. The steady state solution of the Eq.(2) together with Eq.(1) gives

$$I_N = f_N Q S_x(\omega_x) = f_N Q S_0 \frac{(\omega_x - \omega)\omega_{eff}}{(\omega_x - \omega)^2 + \omega_{eff}^2} = f_N Q S_0 \frac{(\omega_{Zeeman} - \omega + AI_N/\hbar)\omega_{eff}}{(\omega_{Zeeman} - \omega + AI_N/\hbar)^2 + \omega_{eff}^2} \quad (C1)$$

By introducing parameters $a = A/\hbar|\omega_{eff}|$ and $b = f_N Q S_0 \omega_{eff}/|\omega_{eff}|$ it can be rewritten as

$$I_N = b\frac{(D + aI_N)}{(D + aI_N)^2 + 1} \quad (C2)$$

where $D = \dfrac{\omega_{Zeeman} - \omega}{|\omega_{eff}|}$ is the relative detuning from the non-perturbed microwave resonance.

Eq.(C2) can be calculated numerically to obtain the dependence $I_N(D)$. The data are shown in Fig.2A under $a = 200$, $b = -0.5$. In a similar way one can calculate the dependence of the modulus $\left|\dfrac{\omega_x - \omega}{\omega_{eff}}\right| = |D + aI_N|$ versus detuning $D = \dfrac{\omega_{Zeeman} - \omega}{|\omega_{eff}|}$ as shown in Fig.2B. There is the region $D \in [-50\mu eV, +50\mu eV]$ where the actual Larmor precession frequency $\omega_x$ is locked to the modulation frequency $\omega$ value. Moreover the dependence is many-valued: there are regions with three solutions. However, not all of them are stable. The question of stability of stationary states can be solved with the help of dynamic equation (2). Indeed, with small deviation of nuclear polarization $I_N'$ from the stationary value $I_N$, the Eq. (2) takes the form $dI_N'/dt = \lambda I_N'$, where Lyapunov exponent

$$\lambda = -\frac{1}{T}\left(1 - f_N Q \frac{\partial S_x(I_N)}{\partial I_N}\bigg|_{\substack{stationary \\ point}}\right) \quad (C3)$$

complies with the stable (unstable) state at $\lambda < 0 (\lambda > 0)$; time $T = \dfrac{T_{1e} T_{1N}}{T_{1e} + T_{1N}}$. Numerical calculation shows that the Lyapunov exponent is increased a few hundred times in the plato region in comparison with the case of the negligible feedback ($f_N = 0$). This enhancement of $\lambda$ indicates a high stability of the given state, when small fluctuations vanish rapidly.

The plato region becomes unstable when the detuning from the optical resonance is negative $(\Delta < 0)$. In this case the parameter $b > 0$. Stationary states of ENSS are shown in Fig.4 (A,B). They are calculated with Eqs. (1,2) for $a = 200, b = +0.5$. Although the bistability regions still exist, the $\omega_x$ locking is absent.

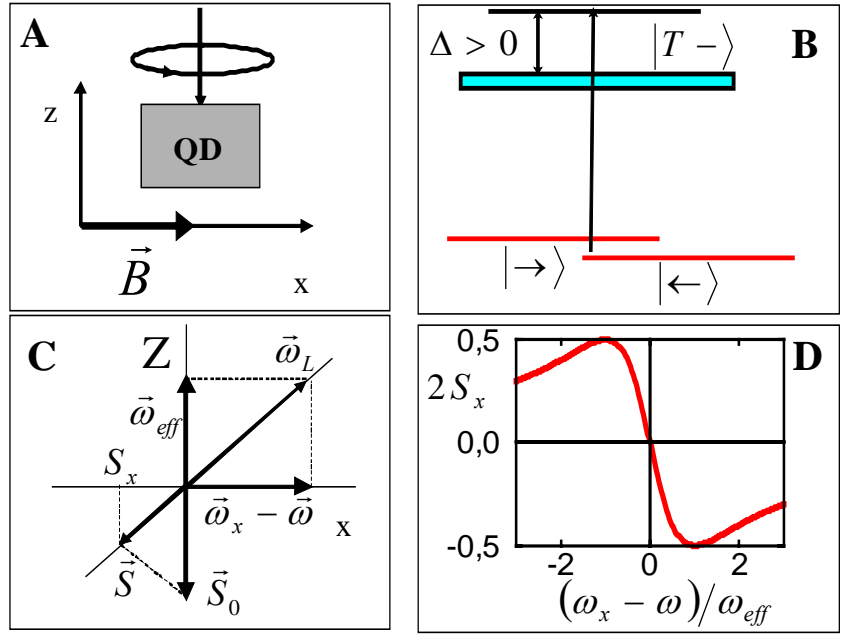

Figure 1 (Color online). (**A**) Scheme of optical pumping of the electron spin in a quantum dot by circularly polarized light (z-axis) in a transverse-to-the-beam magnetic field $\vec{B}$ ($x$-axis). (**B**) Energy diagram of the optical pumping. Ground state has electron spin projections $|\rightarrow\rangle$ and $|\leftarrow\rangle$ onto $x$-axis, $|T-\rangle$ is the excited state of a trion. The orthogonal-to-it state $|T+\rangle$ is not shown. (**C**) Vector diagram of the electron spin precession and the steady state mean electron spin in the rotating frame. (**D**) The dependence of the $2 \cdot S_x$ spin projection on the detuning $\omega_x - \omega$.

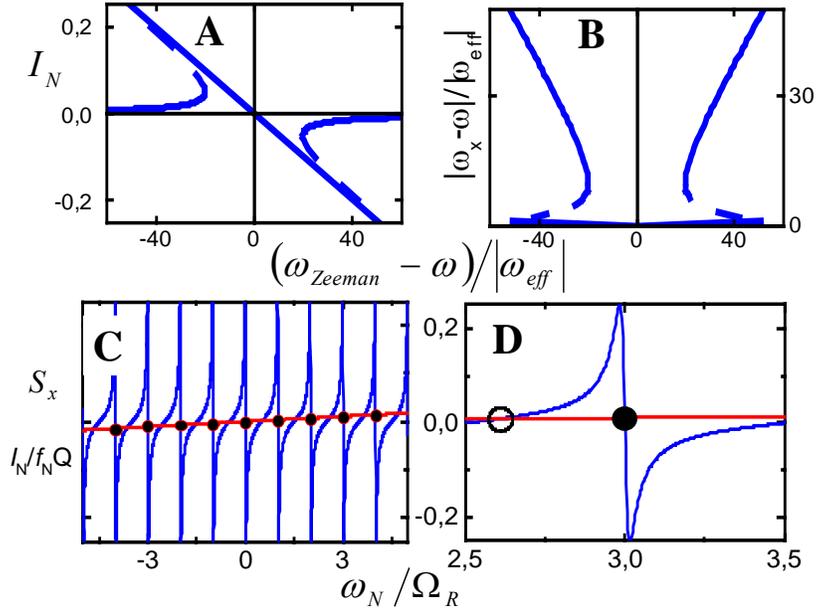

Figure 2 (Color online). Stationary states of ENSS at single frequency modulation (A,B) and pulsed excitation (C,D). (**A**) Nuclear spin $I_N$ vs detuning $\omega_{Zeeman} - \omega$ from non-perturbed resonance. Solid (dashed) lines show stable (unstable) states. (**B**) The dependence of spin precession frequency $\omega_x$ in total external and Overhauser fields (relative to the modulation frequency $\omega$) vs detuning $\omega_{Zeeman} - \omega$. Plato demonstrates the nuclear-assisted frequency locking $\omega_x \approx \omega$. (**C**) The fragment shows 20 steady states of the mean nuclear spin in a quantum dot. They are given by the points of crossing of the red straight line with the blue curve. Solid points show stable states. Actual number of states mounts to a few hundred. (**D**) The magnified (C) region near $\omega_N = 3\Omega_R$. In each pair of crossing points only solid points are stable.

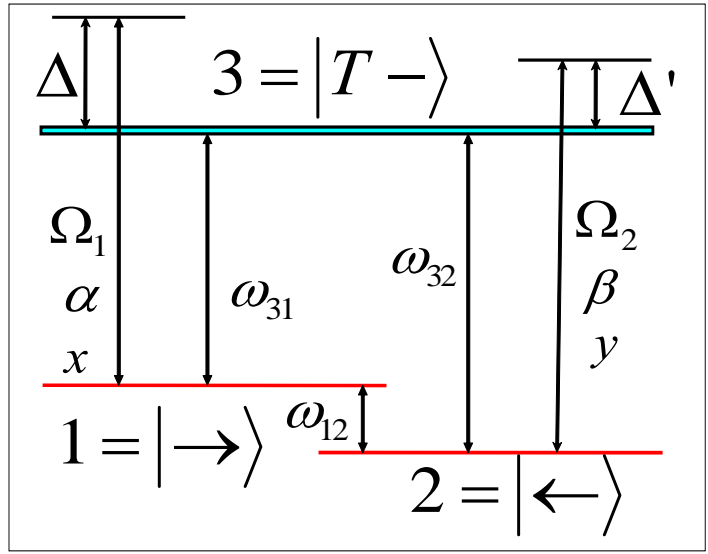

Figure 3 (Color online). The $\Lambda$-scheme for the case of the two-color coherent excitation with $x-$ and $y-$ polarized lasers with frequencies $\Omega_1$ and $\Omega_2$ and Rabi frequencies $\alpha$ and $\beta$, respectively.

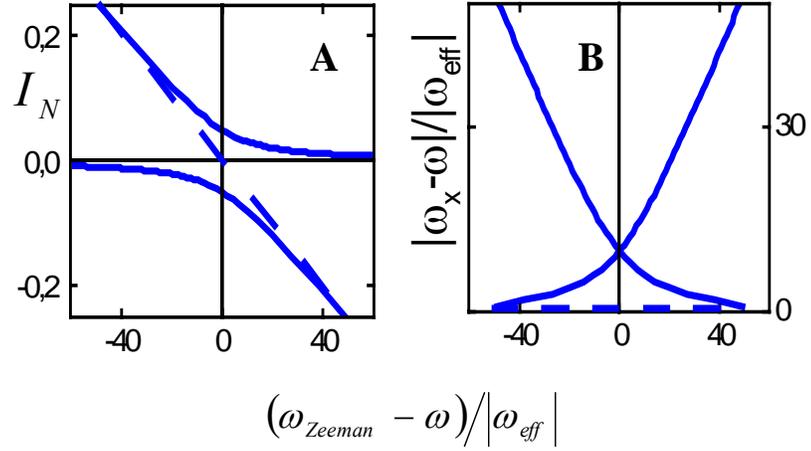

Figure 4 (Color online). Stationary states of ENSS at single frequency modulation (A,B) when detuning from the optical resonance is negative $(\Delta < 0)$ calculated for $a = 200, b = +0.5$. (**A**) Nuclear spin $I_N$ vs detuning $\omega_{Zeeman} - \omega$ from non-perturbed resonance. Solid (dashed) lines show stable (unstable) states. (**B**) The dependence of spin precession frequency $\omega_x$ in total external and Overhauser fields (relative to the modulation frequency $\omega$) vs detuning $\omega_{Zeeman} - \omega$. Plato region near $\omega_x \approx \omega$ is unstable at $b > 0$.

nuclear ensemble. In each of them the fluctuations are strongly suppressed ("narrowed distribution") due to the feedback.

26. At first sight, the frequency $\omega_N$ is already digitalized, since the nuclear spin projection onto $\vec{B}$ changes with $1/N$ steps. However the spacing $AI/\hbar N$ is 25 times less than $\Omega_R$ value, so that the frequency $\omega_N$ and the spin $I_N$ can be considered as continuous variables.

27. One might want to replace the Eq.(2) for the average spin $I_N$ with the equation for the distribution function for the nuclear polarization to take into account the fluctuations $\Delta\omega_N \geq \Omega_R$. However, strong feedback effectively suppresses the Overhauser field fluctuations near the stable stationary states (the so-called "narrowed distribution"). The absolute value of Lyapunov exponent $|\lambda|$ increases by a few orders of magnitude in these states. For example, in the case of pulsed excitation the enhancement is $\sim A/\hbar\Omega_R \sim 300$. Therefore the same driving random force produces 300-times decreased response of both $\Delta I_N$ and $\Delta\omega_N$. As a result the fluctuations of $\omega_N$ in stable points are much less than the intermode distance $\Omega_R$. In these conditions one can safely use the Eq.(2) for the mean nuclear spin.

28. The non-polarized quantum dot ensemble is broken into a few modes of precession due to the nuclear spin fluctuations. Since the number of nuclei $N \sim 10^4$ is finite, there are fluctuations of nuclear spin $I_N \sim I/\sqrt{N}$ and the corresponding scattering of frequencies $\Delta\omega_N \sim AI/\hbar\sqrt{N}$ from dot to dot. This gives a small number of modes of precession $\Delta\omega_N/\Omega_R \sim 5$ as suggested in Refs.[5, 8]. In contrast to it, the spin-polarized ensemble (scattered along the tilted straight line in Fig.2A) will be distributed over the much larger number (>100) of modes (Fig.2C) after pulsed excitation.

29. Although sharing the opinion of Ref.[5] the Ref.[8] made a conclusion about the increased (decreased) stability of a few modes for $\Delta > 0$ $(\Delta < 0)$, thus making an important step toward understanding the nuclear field locking phenomenon. As I discussed here the actual number of modes in a single dot is much larger.